# Effective electro-optical modulation with high extinction ratio by a graphene-silicon microring resonator

Yunhong Ding*(1), Xiaolong Zhu(1,2), Sanshui Xiao(1,2), Hao Hu(1), Lars Hagedorn Frandsen(1), N. Asger Mortensen(1,2), Kresten Yvind (1)

(1) Department of Photonics Engineering, Technical University of Denmark, DK-2800 Kongens Lyngby, Denmark
(2) Center for Nanostructured Graphene, Technical University of Denmark, DK-2800 Kongens Lyngby, Denmark

yudin@fotonik.dtu.dk

**Abstract:** Graphene opens up for novel optoelectronic applications thanks to its high carrier mobility, ultra-large absorption bandwidth, and extremely fast material response. In particular, the opportunity to control optoelectronic properties through tuning of Fermi level enables electro-optical modulation, optical-optical switching, and other optoelectronics applications. However, achieving a high modulation depth remains a challenge because of the modest graphene-light interaction in the graphene-silicon devices, typically, utilizing only a monolayer or few layers of graphene. Here, we comprehensively study the interaction between graphene and a microring resonator, and its influence on the optical modulation depth. We demonstrate graphene-silicon microring devices showing a high modulation depth of 12.5 dB with a relatively low bias voltage of 8.8 V. On-off electro-optical switching with an extinction ratio of 3.8 dB is successfully demonstrated by applying a square-waveform with a 4 V peak-to-peak voltage.

In addition to holding novel electronic properties, graphene is now also emerging as a material of interest in the area of optoelectronics. Graphene has many unique properties, such as zero-band gap and tunable Fermi level [1, 2], ultra-broad absorption bandwidth [3, 4], high carrier mobility around 200000 $cm^2V^{-1}s^{-1}$ at room temperature [5, 6], and a super high Kerr coefficient for high nonlinearity applications [7, 8]. Those interesting properties give rise to many potential applications [9, 10], such as wafer-scale integrated circuits [11], solar cells [12, 13], high-speed graphene-silicon electro-optical modulators [14-16], optical-optical switches [17, 18], saturation absorbers [19-21], photodetectors [22-24], and nonlinear media for four-wave mixing (FWM) [25, 26].

The deployment of graphene on top of a silicon waveguide is an efficient mean to make graphene-silicon hybrid devices. In order to electrostatically tune the Fermi level of graphene, there is a need to sandwich a thin layer of material with a high dielectric constant (e.g. $Al_2O_3$ [27, 28] or $Si_3N_4$ [29]) between the silicon layer and the added layer of graphene. When this graphene-silicon capacitor is biased, carriers can be either accumulated on the graphene sheet, or swept out from the graphene sheet, resulting in a convenient tuning of the Fermi level and, thus, optical absorption [14]. This technology has enabled high-speed electro-optical modulators [14-16]. However, the strong light confinement in the high-index silicon gives a modest optical field overlap with the graphene layer. Hence, the graphene-light interaction is consequently too low to obtain a significant modulation depth. In order to enhance the modulation depth, a graphene-silicon microring resonator has recently been used to reach a modulation depth of 40% (2.2 dB) [30].

In this paper, we comprehensively study the interaction between graphene and a microring resonator, and its influence on modulation depth and graphene-coverage tolerance. Based on the theoretical analysis, we specially design and demonstrate graphene-silicon microring resonators. By designing the microring for operation in the slightly under-coupling while being close to the critical-coupling region, a high modulation depth of 12.5 dB is demonstrated using a relatively low bias voltage of 8.8 V. Increasing the graphene length results in a slight detuning from the critical-coupling condition and, thus, a lower modulation depth of 6.8 dB is achieved. An on-off electrical switching signal with a 4 V peak-to-peak voltage is further applied on the graphene-silicon microring resonator, and clear optical on-off switching with an extinction ratio of 3.8 dB is demonstrated.

**Tunability of graphene-silicon waveguide**

Tuning the resonance of a graphene-silicon microring resonator is based on a tuning of the waveguide loss, therefore, the tunability of a graphene-silicon straight waveguide is firstly investigated. Fig. 1 shows the three dimensional (3D) schematic of a graphene-silicon waveguide. One side of the 450 nm wide silicon waveguide is shallowly etched while the other side is fully etched. The gold (Au) contact for silicon is designed on top of the shallowly etched silicon layer. A thin $Al_2O_3$ layer is introduced between the silicon layer and the graphene cover. A gold/chromium (Au/Cr) contact is used for electrical contacts with graphene thanks to the good contact between Cr and graphene [31]. The Au contacts for silicon and Au/Cr contacts for graphene are both designed to be 3 µm away from the waveguide to avoid absorption loss associated with the metal. Fig. 1(b) illustrates the cross-section of the designed waveguide and the corresponding transverse electric (TE) mode-field profile is shown in Fig. 1(c). The propagation loss of the designed graphene-silicon waveguide, as presented in Fig. 1(d), is calculated using the Fermi level dependent complex dielectric constant for graphene [32]. We find that propagation loss decreases as the Fermi level increases, and the largest propagation loss of ~0.1 dB/µm is predicted at the Dirac point. Fig. 1(e) shows a scanning electron micrograph (SEM) image of a fabricated graphene-silicon waveguide where a very good coverage with graphene is found. In order to characterize the quality of the transferred chemical-vapor deposition (CVD) graphene, Raman spectroscopy is employed, since it is able to explore the properties and the structures of graphene, such as the number of layers, doping (Fermi) levels, and dopants or defects [33, 34]. Fig. 1(f) shows the measured Raman spectrum (excited at 532 nm) of the transferred graphene. Prominent features with Raman shifts of the G-peak around 1585 $cm^{-1}$ and the 2D-peak around 2683 $cm^{-1}$ are observed. The single Lorentzian shape of the 2D-peak and the height ratio between the G- and 2D-peaks indicate the signature of a monolayer graphene. The low-intensity peak around 1345 $cm^{-1}$ is attributed to the D-band of carbon, which is activated by the presence of defects in the transferred CVD graphene, such as vacancies or grain boundaries, as well as adsorbents remaining from the graphene transfer process.

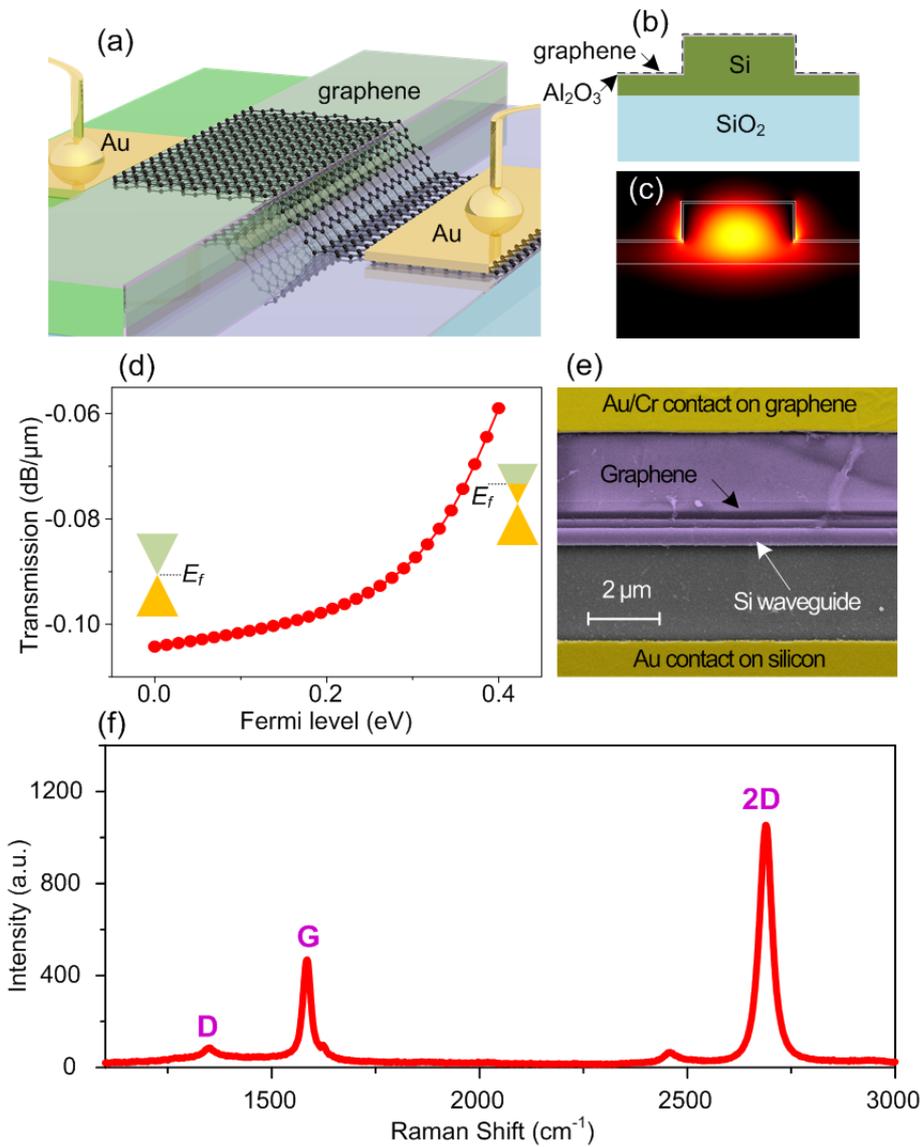

**Figure 1.** (a) 3D schematic of the designed graphene-silicon waveguide. (b) The cross-section of the waveguide, and (c) the corresponding TE mode-field profile. (d) Calculated transmission loss of the designed graphene-silicon waveguide as a function of the Fermi level of graphene. (e) False-color SEM top-view image of the fabricated graphene-silicon waveguide. (f) Raman spectrum of the transferred graphene sheet.

In order to investigate the propagation loss, graphene-silicon waveguides with different graphene interaction lengths were fabricated and measured. Photonic crystal grating couplers [35] are used for coupling light in and out of the devices. The obtained spectra are shown in Fig. 2(a), and a linear propagation loss α of 0.07 dB/μm is obtained by the cut-back method, as presented in Fig. 2(b), which is in agreement with the calculations. Note that the CVD graphene is normally p-doped [36], thus, a lower loss than the case with a Fermi level of 0 eV is expected. The transmitted optical powers for graphene lengths of 500 μm and 800 μm deviate from the linear fit because of imperfect coverage of the graphene by crack areas and are, thus, not part of the fit. The tunability of the graphene-silicon waveguide is further explored by applying different bias voltages on the graphene layer as indicated in Fig. 3. We find that the propagation loss of the graphene-silicon waveguide is effectively tuned by applying different voltages. When the bias voltage is low (-1 V to 6 V), the Fermi level $E_f$ is close to the Dirac point ($E_f<h\nu_0/2$). Photons

(with energy $h\nu_0$) propagating through the waveguide experience a high damping due to inter-band transitions in the graphene. When a higher negative voltage (<-1 V) is applied, the Fermi level is lowered (<-$h\nu_0/2$), and light propagation is consequently less attenuated since no electrons are available for inter-band transitions. On the other hand, a high positive voltage (>6 V) results in a higher Fermi level (>$h\nu_0/2$). In this case, a lower optical propagation loss is also obtained since all electron states are filled up in graphene and no inter-band transition is allowed due to Pauli blocking. The propagation loss tunability of 0.04 dB/µm (from 0.07 dB/µm to 0.03 dB/µm) is achieved when the graphene is negatively biased.

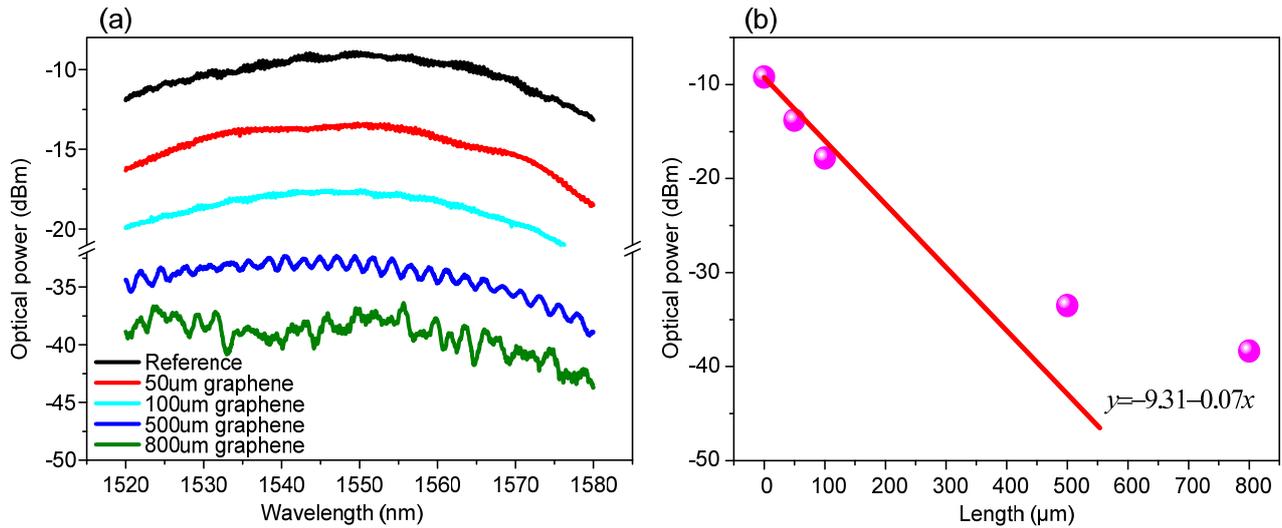

Figure 2. (a) Measured transmitted optical power of the graphene-silicon waveguide as a function of wavelength for different graphene lengths. (b) Cut-back measurements of the graphene-silicon waveguide at wavelength of 1555 nm.

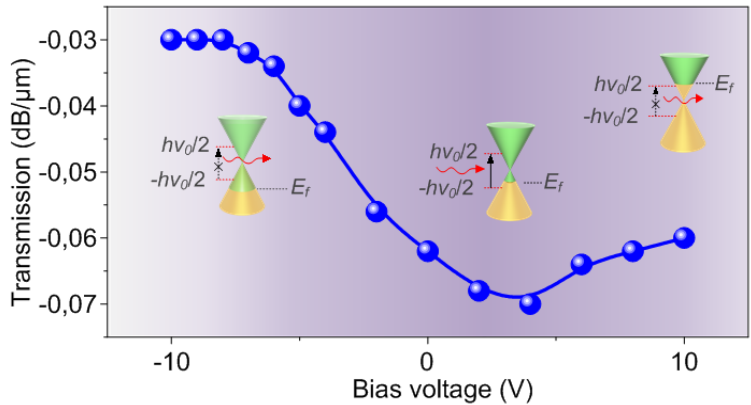

Figure 3. Measured transmission loss of the graphene-silicon waveguide as a function of bias voltage.

**Graphene-silicon microring resonator**

In order to obtain a high enhancement of intensity modulation, it is important to understand the interaction between graphene and microring resonator. An all-pass type silicon microring resonator with radius of 50 µm incorporating with a graphene layer on part of the ring waveguide is analyzed. Fig. 4(a) shows the modulation depth (defined as $ER_2-ER_1$, where $ER_1$ and $ER_2$ are the extinction ratio before and after tuning graphene) as a function of power coupling coefficient $\kappa^2$ of the microring resonator and graphene coverage on the ring waveguide. Here typical linear propagation loss for pure silicon and graphene-silicon waveguides are considered to be 2 dB/cm and 0.1 dB/μm respectively, and propagation loss tunability $\Delta\alpha$ of 0.02 dB/μm for graphene-silicon waveguide is considered. We find for each power coupling coefficient, there are two optimum graphene coverages, which are corresponding to critical-coupling, i.e. $a = r$, for before (dark blue region) and after (dark red region) tuning of graphene, respectively. $a$ is the roundtrip-transmission coefficient, and $r = \sqrt{1 - \kappa^2}$. Note that critical-coupling of graphene-silicon microring after tuning graphene also means under-coupling ($r>a$) before tuning graphene. From Fig. 4(a), we also find that increasing $\kappa^2$ results in increasing of both two optimum coverage values. The detailed dependency of modulation depth to graphene coverage for $\kappa^2 = 0.75$ (the red dashed line in Fig. 4(a)) is exhibited in Fig. 4(b). The negative extreme-value (working point A, corresponding to 19% coverage by graphene) indicates that initially graphene-silicon microring is working in critical-coupling condition with high extinction ratio (ER). In this case, tuning graphene makes the microring resonator detuned from critical-coupling with lower ER, thus, large modulation depth is obtained, as shown in Fig. 4(c). The positive extreme-value (working point B, corresponding to 24% coverage by graphene) indicates that initially graphene-silicon microring resonator is working in under-coupling ($r>a$) with low ER. In this case, tuning graphene reduces the propagation loss leading to a larger $a$, thus microring resonator is tuned to work in critical-coupling with higher ER, and large modulation depth can also be obtained, as indicated in Fig. 4(d). In real fabrication process, there can be broken area of graphene which leads to an imperfect coverage, and coverage tolerance for a required modulation depth is important. Fig. 4(e) shows the coverage tolerance that guarantees >10 dB absolute modulation depth ($|ER_2-ER_1|$) as a function of power coupling coefficient $\kappa^2$ for different propagation loss tunability of graphene-silicon waveguide. Firstly, we find that if graphene is working on critical-coupling after tuning graphene (i.e. under-coupling before tuning graphene), a larger coverage tolerance is obtained. In addition, both larger power coupling coefficient and larger propagation loss tunability result in a larger coverage tolerance. Thus, in practice, it is preferable to design the graphene-silicon microring resonator to have large power coupling coefficient with proper coverage, so that initially it works in slightly under-coupling condition and in critical coupling after tuning graphene.

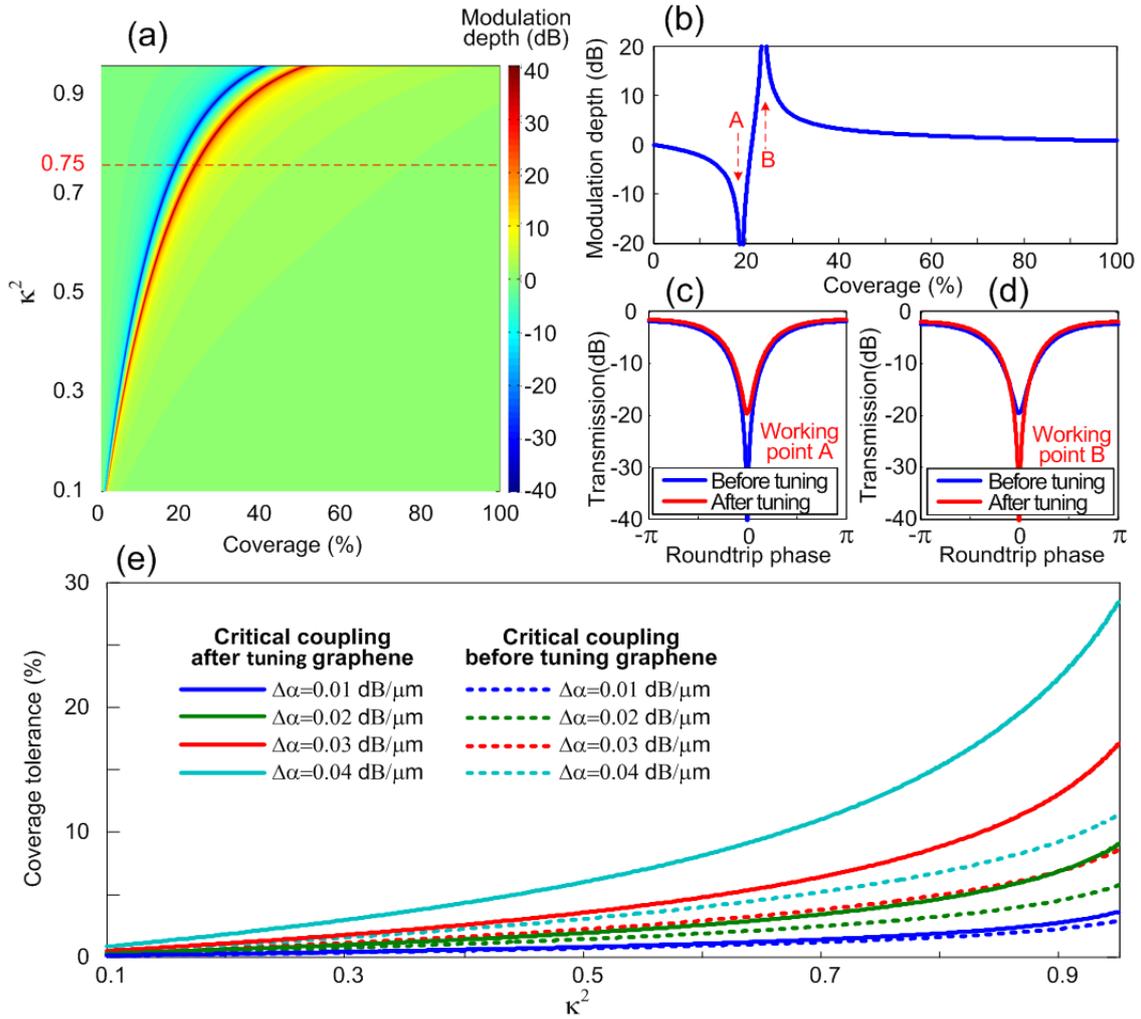

**Figure 4.** (a) Modulation depth as a function of power coupling coefficient $\kappa^2$ and graphene coverage on the ring waveguide, $\Delta\alpha$=0.02 dB/μm, and (b) detailed modulation depth as a function of graphene coverage with $\kappa^2$=0.75 (the red dashed line in (a)). Corresponding transmissions of graphene-silicon microring at (c) working point A with graphene coverage of 19%, and (d) point B with graphene coverage of 24%. (e) Graphene coverage tolerance (absolute modulation depth >10 dB) as a function of power coupling coefficient of the microring for different propagation loss tunability of the graphene-silicon waveguide.

Based on the analysis describe above, an all-pass type graphene-silicon microring resonator, as schematically shown in Fig. 5(a), is designed with a high power-coupling coefficient $\kappa^2$ of 0.75. According to previous analysis, an optimum graphene coverage of 25% on the ring waveguide is selected, so that a 10 dB modulation depth can be guaranteed with a large coverage tolerance of 13% (i.e. ±6.5%) for a propagation loss tunability of 0.04 dB/μm. Fig. 5(b) exhibits the SEM image of the fabricated device, indicating a quarter-coverage of graphene on the microring waveguide. A zoom-in of the bended silicon waveguide covered by graphene layer is shown in Fig. 5(c) and reveals again a good coverage by graphene on the silicon waveguide. The width of the bending silicon waveguide is 450 nm. The small coupling gap is designed to be 150 nm, which is corresponding to a required power-coupling coefficient $\kappa^2$ of 0.74 calculated by 3D full vectorial mode-matching method with coupled-mode theory [37].

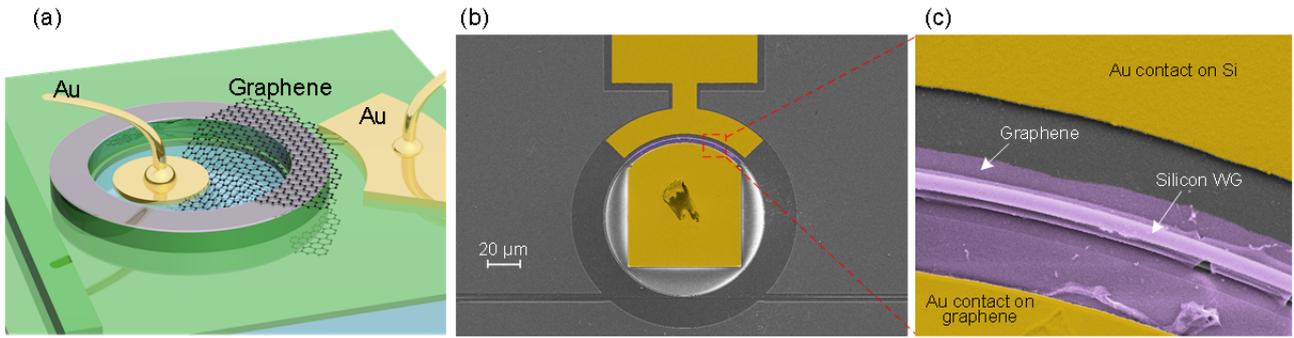

**Figure 5.** (a) 3D schematic image of the graphene-silicon microring resonator. (b) False-color SEM image of the fabricated graphene-silicon microring device and (c) a zoom-in of the bended silicon waveguide covered by graphene.

The tunability of the graphene-silicon microring resonator (with 25% graphene coverage) is firstly investigated by applying different bias voltages to the graphene. For zero bias (black, Fig. 6(a)), a high extinction ratio of 25.5 dB is achieved, indicating that graphene is working close to the critical-coupling region. From Fig. 3, we know that a higher tuning range can be obtained for the graphene-silicon waveguide in the negative-bias region. With a stronger negative bias, the extinction ratio of the microring transmission increases. For a bias of -8.8 V, we obtain a high extinction ratio of 38 dB. Furthermore, we see a slight blue shift of the resonance, which is an effect of a bias-induced reduction in the effective refractive index of the graphene-silicon waveguide [30, 32]. Fig. 6(b) shows the measured modulation depth as a function wavelength for different bias voltages. High modulation depth is obtained around the resonance wavelength, and the highest modulation depth of 12.5 dB is found for a bias of -8.8 V. The narrow bandwidth of the modulation depth is inherently due to the high Q-value of the ring resonator. The measured transmission without bias is further fitted, and a roundtrip transmission coefficient $a$ of 0.45 and a power-coupling coefficient $\kappa^2$ of 0.77 (corresponding to $r$=0.48) are found, indicating that the microring resonator is in slightly under-coupling condition as designed. With $a$ and $\kappa^2$, the dependency of the transmission to a change in propagation-loss ($\Delta\alpha$) is investigated, as shown in Fig. 6(c). Indeed, since the microring resonator is working close to critical-coupling, a modest $\Delta\alpha$ of 0.006 dB/µm will cause a large transmission change over 12 dB for the graphene-silicon microring resonator. Note that -8.8 V is expected to achieve a higher propagation loss tunability of 0.03 dB/µm as indicated in Fig. 3. The lower tunability of 0.006 dB/µm as used in the simulation is due to potential crack areas of the graphene coverage.

In order to compare the effect of different graphene coverage lengths, graphene-silicon microring resonators with half and three-quarter coverages by graphene are fabricated and measured, as shown in Fig. 5(d) and Fig. 5(g), respectively. We find that compared with the microring with quarter-coverage, a larger negative bias voltage of -12.5 V is needed to generate a significant transmission change for half coverage by graphene, since now the graphene-microring resonator is detuned far away from the critical-coupling condition. At the same time, since larger bias voltage is applied, a stronger blue shift of the resonance is obtained. The corresponding modulation depth as a function of wavelength with different bias voltages is presented in Fig. 5(e). The largest modulation depth of 6.2 dB is obtained for a bias voltage of -12.5 V. A wide bandwidth is also achieved which is due to the fact that a longer segment covered by graphene results in a lower Q-value. With the same power-coupling coefficient $\kappa^2$ of 0.77 and a roundtrip transmission coefficient $a$ of 0.28, the measured transmission without bias is also fitted. Again, the dependency of the transmission on the change in propagation loss $\Delta\alpha$ is analyzed and shown in Fig. 5(f). A larger change in propagation loss is needed to ensure a significant transmission change, which agrees with the theoretical analysis and confirms the experimental measurements. Compared with the device with half-coverage by graphene, the microring with three-quarter coverage by graphene exhibits slightly higher modulation depth with narrower bandwidth for the same bias voltage, as shown in Fig. 6(h). A maximum 6.8 dB modulation depth is obtained with a bias voltage of -12.5 V. This is due to imperfect graphene coverage, which results in an actual shorter coverage than the device with half coverage by graphene. Indeed, we find a fitting parameter of *a*=0.3, which is higher than that of the device with half-coverage by graphene, indicating a shorter actual-coverage, as analyzed in Fig. 6(i).

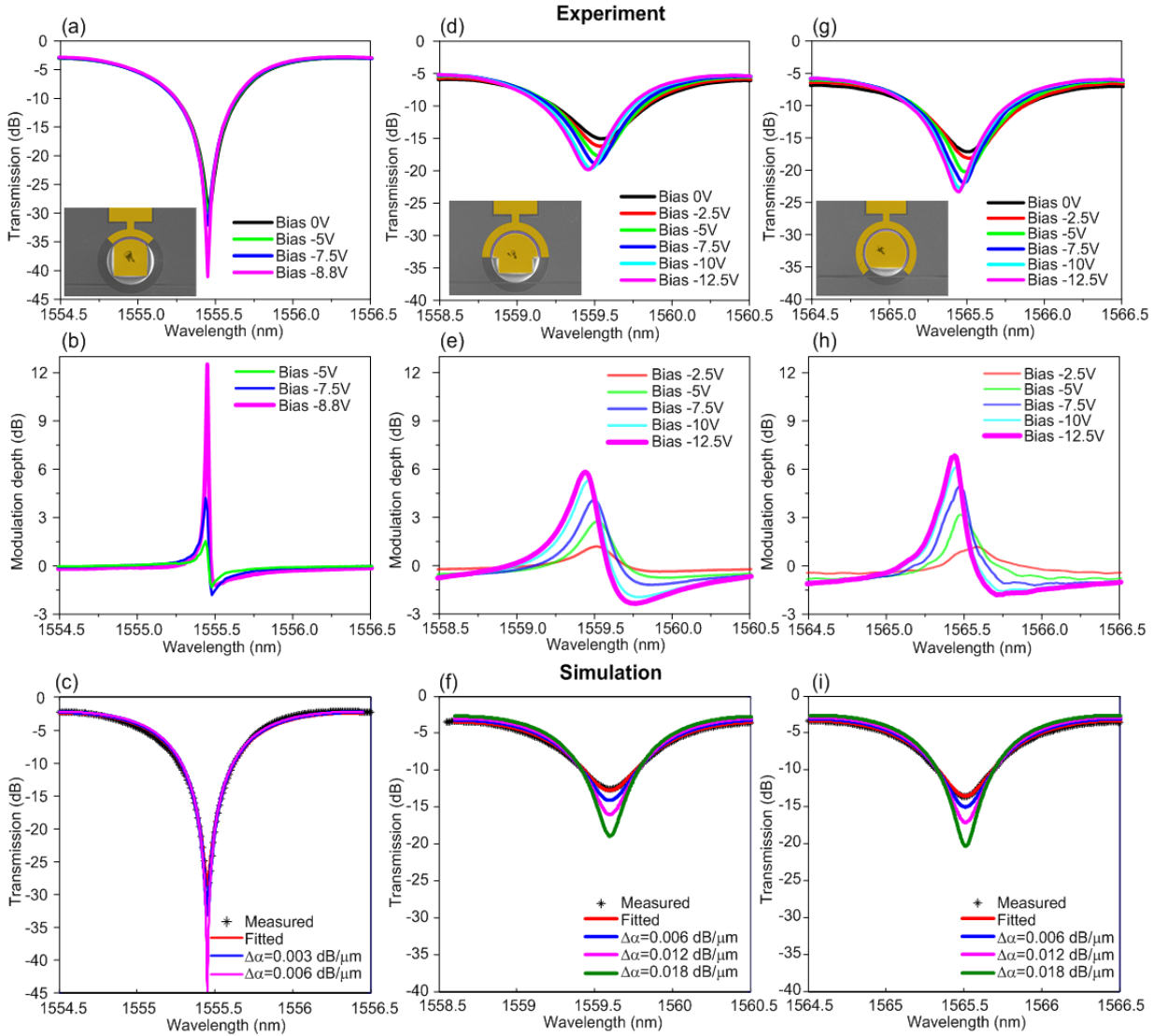

**Figure 6.** (a) Measured transmission and (b) the corresponding modulation depth of graphene-silicon microring with quarter-coverage by graphene for different bias voltage on graphene. The inset in (a) shows the fabricated device. (c) Theoretical analysis of the graphene-silicon microring with quarter- coverage for different change of absorption coefficient of the graphene-silicon waveguide. $\kappa^2$=0.77, *a*=0.45. (d) Measured transmission and (e) the corresponding modulation depth of graphene-silicon microring with half coverage by graphene for different bias voltage on graphene. The inset in (d) shows the fabricated device. (f) Theoretical analysis of the graphene-silicon microring with half coverage for different absorption coefficient change of the graphene-silicon waveguide. $\kappa^2$=0.77, *a*=0.28. (g) Measured transmission and (h) the corresponding modulation depth of graphene-silicon microring with three-quarter coverage by graphene for different bias voltage on graphene. The inset in (g)

exhibits the fabricated device. (i) Theoretical analysis of the graphene-silicon microring with three-quarter coverage for different absorption coefficient change of the graphene-silicon waveguide. $\kappa^2$=0.77, *a*=0.3.

Considering the graphene-silicon microring with three-quarter coverage by graphene shows good modulation depth with relative larger bandwidth, it is further tested by applying a square waveform bias in order to explore the dynamic switching, as shown in Fig. 7. The wavelength is selected to be 1565.45 nm where the graphene-silicon microring resonator is on-resonance when negatively biased. It can be seen that the square waveform with a peak-to-peak voltage of 4 V, Fig. 7(a), results in an effective optical switch with an extinction ratio of 3.8 dB, as shown in Fig. 7(b). It should be noted that our device is not designed for high-frequency operation, which would need further optimization of electrical contacts on the device in order to reduce the associated resistive-capacitive (RC) time constant of the RC circuit. Reducing the contact resistance as well as capacitance allows the devices to work over GHz [14-16]. The resistance can be reduced by decreasing the distances from metal-contact pads on graphene and contact pads on silicon to the silicon waveguide. More efficient ohm contact with silicon, i.e. Al/Si contact, can be used in order to further decrease the resistance. The capacitance can be reduced by increasing the $Al_2O_3$ thickness.

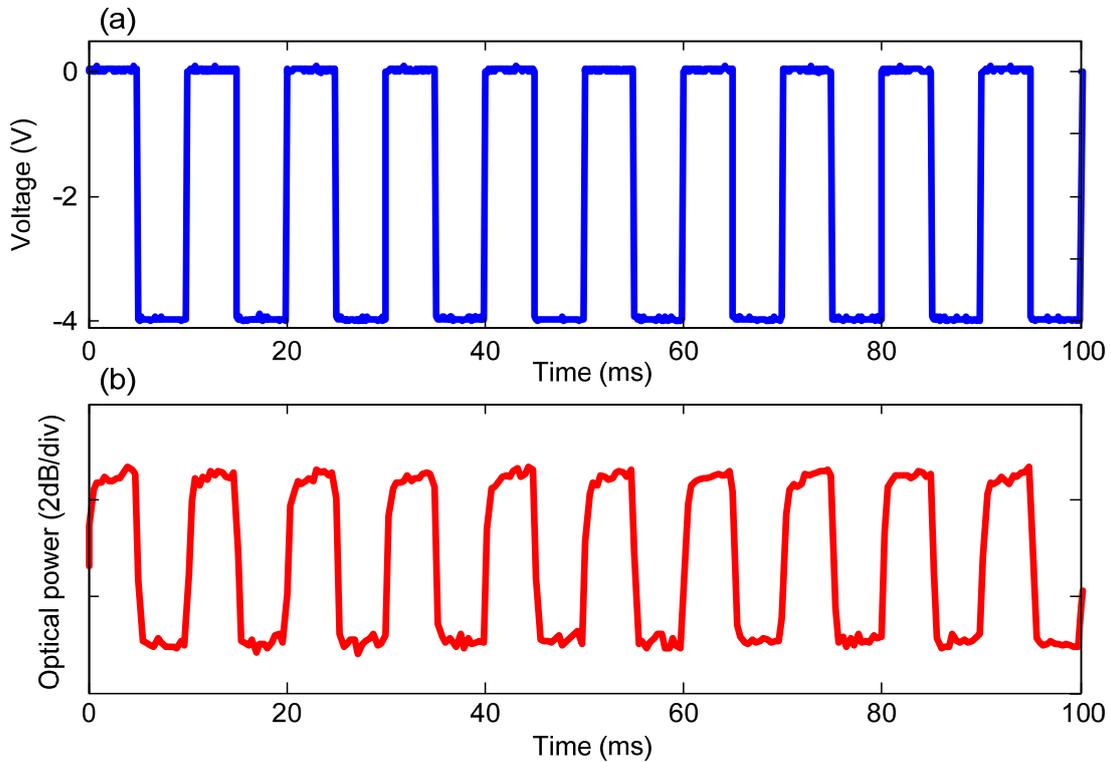

Figure 7. On-off switching of the graphene-silicon microring resonator with three-quarter coverage by graphene. (a) The applied electrical square waveform with a peak-to-peak voltage of 4 V. (b) The measured modulated optical signal with an on-off ratio of 3.8 dB.

**Conclusion**

In conclusion, we have comprehensively analyzed the interaction of graphene and a microring resonator, and its influence on modulation depth and graphene-coverage tolerance. This provides important guidelines for design of graphene-silicon microring resonators. We have further demonstrated a graphene-silicon hybrid microring resonator, and different coverage lengths by graphene are fabricated and compared. A high extinction ratio of 12.8 dB is obtained with relatively low negative bias voltage of 8.8 V on graphene when the silicon microring is covered with optimum coverage length by graphene and working slightly in under-coupling while close to critical-coupling. Higher negative bias is needed to obtain a significant transmission change if the graphene-silicon microring resonator is with longer graphene coverage. Effective electro-optical on-off switching with an extinction ratio of 3.8 dB is obtained by applying a square waveform electrical bias with a peak-to-peak voltage of 4 V on the graphene.

**Method**

The device is fabricated in on a commercially available silicon-on-insulator (SOI) wafer with a top-silicon layer of 250 nm and a buried oxide (BOX) layer of 3 μm. The top-silicon layer is p-type with a resistivity of ~20 Ω·cm. Firstly, standard SOI processing, including e-beam lithography (EBL) and inductively coupled plasma (ICP) etching, was used to fabricate the shallowly etched silicon microring resonators. In order to obtain good optical fiber coupling to the integrated waveguides, 12 μm-wide photonic crystal grating couplers [35] with a 400 μm long adiabatic taper are used to connect the input and output of the bus waveguide of the silicon microring resonator. A second SOI processing is used to fully etch the inside the ring where the graphene contacts will be placed. Then, the contacts on the silicon were defined by standard ultraviolet (UV) lithography and followed by Au deposition and liftoff processing. After that, a ~9 nm thin $Al_2O_3$ layer was grown by atom-layer deposition (ALD) machine on the silicon chip. Then, a 1.5 cm×1.5 cm graphene sheet grown by CVD was wet-transferred [38, 39] on top of the silicon devices. In this step, a polymer film, e.g. poly methyl methacrylate (PMMA), was firstly spin-coated onto the graphene covered copper foil and dried at 170 °C for 1 min. Subsequently, a PMMA/graphene membrane was obtained by etching away the copper foil in a Fe(NO3)3/H2O solution and transferred onto the waveguide or ring resonators devices. Finally, the PMMA was dissolved in acetone that also cleaned the graphene surface. The coverage areas on the ring resonators were then defined by standard UV lithography followed by oxygen (O2) plasma etching. Following that, the contact windows on graphene were defined by UV lithography, and Au/Cr contacts on the graphene were finally obtained by Au/Cr metal deposition and a liftoff process.

**Acknowledgment**

This work is supported by the Danish Council for Independent Research (DFF-1337-00152 and DFF-1335-00771). The Center for Nanostructured Graphene is sponsored by the Danish National Research Foundation, Project DNRF58.